\documentclass[aps,prd,preprint,letterpaper,showpacs,groupedaddress]{revtex4}
\usepackage{natbib}
\usepackage{graphicx}
\usepackage{bm} 

\begin{document}


\newcommand{\MET}{\mbox{$\raisebox{.3ex}{$\not$}E_T$\hspace*{0.5ex}}}
\newcommand{\gevcc}{GeV/$c^{2}$}
\newcommand{\ttbar}{$t\overline{t}$}
\newcommand{\bbbar}{$b\overline{b}$}
\newcommand{\ccbar}{$c\overline{c}$}
\newcommand{\qqbar}{$q\overline{q}$}
\def\ET{$E_{T}$}
\def\pt{$p_{T}$}
\def\MW{M_{W}}
\def\MH{M_{H}}
\def\Mt{m_{t}}
\def\Mb{m_{b}}
\def\micron{\mu{\rm m}}
\def\fb{\rm{fb^{-1}}}
\def\pb{\rm{pb^{-1}}}
\def\Lxy{$L_{xy}$}
\newcommand{\lxybar}{$\langle L_{xy} \rangle$}

\title{
A method for measurement of the top quark mass 
using the mean decay length of $b$ hadrons in \ttbar~events.}


\author{C. S. Hill}
\affiliation{The University of California Santa Barbara}

\author{J. R. Incandela}
\affiliation{The University of California Santa Barbara}

\author{J. M. Lamb}
\affiliation{The University of California Santa Barbara}

\date{\today}
\begin{abstract}

We present a new method for the experimental determination 
of the top quark mass that is based upon the mean distance of travel 
of $b$ hadrons in top quark events. The dominant systematic uncertainties 
of this method are not correlated with those of other methods, but a
large number of events is required to achieve small statistical 
uncertainty. Large \ttbar~event samples are expected from Run II of the 
Fermilab Tevatron and the CERN Large Hadron Collider (LHC). We show that 
by the end of Run II, a single experiment at the Tevatron could achieve a 
top quark mass uncertainty of $\sim$ 5~\gevcc~by this method alone. 
At the CERN LHC, this method could be comparable to all others methods, 
which are expected to achieve an uncertainty of $\sim$1.5~\gevcc~
per experiment. 
This new method would provide a useful cross-check to other methods, and 
could be combined with them to obtain a substantially reduced overall 
uncertainty. 

\end{abstract}


\maketitle

\section{\label{sec:introduction}Introduction}

There are important reasons for measuring the properties of top 
quarks \cite{Chakraborty:2003iw}. If there are new strong interactions 
at the TeV scale, the top quark often plays a central role in the 
corresponding theories. The large top quark mass is important in 
extensions of the standard model, particularly when mass dependent 
couplings are considered. New particles are experimentally constrained 
to be heavier than other fermions but could be lighter than the top quark. 
Careful studies of top quark decay products could therefore be fruitful. Indeed the 
large top quark mass and its approximately unit value Yukawa coupling to the Higgs 
particle are often cited as a hint that the top quark has a special place 
in the greater scheme of things \cite{Quigg:1997uh}.

The mass of the top quark, $\Mt$, is an important Standard Model parameter 
that enters quadratically into higher order corrections to the $W$ mass, 
$\MW$, along with the Standard Model Higgs mass, $\MH$, which enters 
logarithmically. Precise values of both $\MW$ and $\Mt$ can thus be used to 
constrain the Standard Model expectation for $\MH$ \cite{Marciano:2004hb}. 
Prior to observation of the Higgs, this information could be used to guide 
experimental searches. After the Higgs has been observed, it can be used as
an interesting consistency check. If the observed Higgs mass were to be  
inconsistent with the prediction obtained from precision 
measurements of $\Mt$ and $\MW$, 
this would indicate that nature is not governed by a Standard 
Model Higgs sector. For these reasons it is essential to measure $\Mt$ and 
$\MW$ to the best possible precision.

In this article we present a new method for the measurement of the top 
quark mass that is based upon the observation that the mass is correlated 
in a relatively strong and unambiguous way with the mean distance of travel 
of the $b$ hadrons  \cite{lepton} formed from the $b$ quarks produced in top quark decays
\cite{incandela}.  The dominant systematic uncertainties of this method 
are associated with those factors that influence the mean $b$ hadron 
decay length, such as $b$ quark fragmentation, $b$ hadron lifetimes, and our 
ability to understand and to accurately model the momenta of the top quarks 
and their decay products. It will be seen that jet energy scales, 
the dominant systematic uncertainties in other methods, result in a relatively 
small contribution to the uncertainty in the mean $b$ hadron decay length. As 
a result, this technique can provide an independent measurement that can be 
used to cross-check, or be combined with, results from other methods to 
yield a more robust, accurate, and precise final value for $\Mt$.

The top quark was first observed nearly a decade ago in collisions of
protons with anti-protons at a center of mass energy of $\sqrt{s}=1.8$ TeV
during Run I of the Fermilab Tevatron by the CDF and D0 
experiments \cite{Abe:1995hr,Abachi:1995iq}. 
Using the relatively small Run I \ttbar~event 
samples, and combining the results of both experiments, the mass of the top 
quark was determined to be \cite{Azzi:2004rc}:
\begin{equation}
\Mt=178.0\pm 2.7~(stat) \pm 3.3~(sys)~{\rm GeV/c}^2 
\end{equation}
This result was the fruit of a very substantial effort by large teams of
physicists on both experiments that benefited from an innovative 
improvement in methodology \cite{Abazov:2004cs}. 
After Run I, there were substantial upgrades to 
the Fermilab accelerator complex, including an increase of beam energies to
$\sqrt{s}=1.96$ TeV, and to the CDF and D0 experiments. Tevatron Run II began in 
2001 and is scheduled to last until slightly beyond the startup of the CERN 
Large Hadron Collider (LHC). In this period, Tevatron experiments could 
collect top quark samples that are almost two orders of magnitude larger 
than those of Run I. These will enable the CDF and D0 experiments to 
significantly improve their measurements of $\Mt$. Each experiment currently 
expects to achieve an overall uncertainty of roughly 2-3 \gevcc~
\cite{Amidei:1996dt}.

The CERN LHC will supersede the Tevatron as the 
world's highest energy accelerator when it begins operation later this 
decade. The LHC will collide protons with protons at a center of mass 
energy of $\sqrt{s}=14$ TeV. The 7-fold increase in energy relative to 
the Tevatron means that the dominant \ttbar~production mechanism will 
change from quark-anti-quark annihilation to gluon fusion. As a result, 
the \ttbar~production cross-section rises from a calculated value 
of $6.7$~pb \cite{topxsect1,topxsect2} at the Tevatron, to $833$~pb 
\cite{topxsect1} at the LHC. The LHC will also operate at higher luminosity 
than the Tevatron. The net effect is that the LHC will produce $\sim$8M 
\ttbar~events per year \cite{mtopatlas} at its initial operating luminosity 
of 10$^{33}~$cm$^{-2}$~sec$^{-1}$. This number will rise by an order of magnitude 
for high luminosity operation. Availability of such large top quark samples 
will mean that all measurement techniques will rapidly become limited by
systematic uncertainties, which are expected to be 
$\sim1.5$~\gevcc. This will also be true for the extraction of the top quark
mass from the mean $b$ hadron decay length. 

\begin{figure}
\includegraphics[width = 0.8\textwidth]{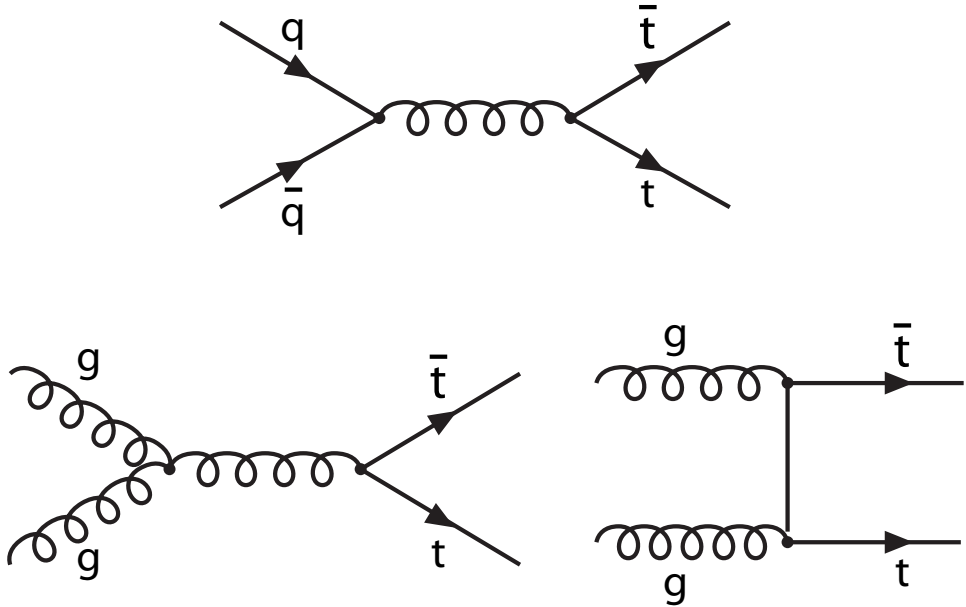}
\caption{\label{fig:tt_prod}Feynman diagrams showing the leading order 
\ttbar~production mechanisms.}
\end{figure}

In this article we present estimates for the potential top quark mass resolution 
of this new method at both the Tevatron and the LHC. In Section 
\ref{sec:method} we describe in detail the aspects of the method that are
common to its application at both accelerators. In particular, we discuss 
the important sources of uncertainty and how they are estimated. The 
approaches taken for the Tevatron and LHC are not identical, mainly due to 
the fact that Tevatron measurements will be statistically limited, 
even for the most optimistic projections of integrated luminosity. On 
the other hand, the large \ttbar~event rates at the LHC allow one to 
quickly achieve statistical uncertainties below 1~\gevcc, even in the 
most pessimistic operating scenarios. This allows us to define event selection
criteria that minimize uncertainties. Descriptions of the 
event selection and several other aspects of our studies that are unique 
to a given accelerator environment are postponed until sections  
\ref{sec:tevatron} and \ref{sec:lhc}, where we also report the results 
of our studies for the Tevatron and LHC, respectively. 
It will be seen that variations in the estimated uncertainties in 
the two accelerator environments can be traced, among other things, to 
a difference in the dominant \ttbar~production mechanism.

Our basic approach is to use Monte Carlo programs \cite{pythia,herwig} 
currently in use by Tevatron and LHC experimental groups to generate 
\ttbar~events at the appropriate beam energies. We do not perform full 
simulations of any of the Tevatron or LHC experiments. Rather, we 
include the effects of imperfect pattern recognition and detector resolution 
by smearing Monte Carlo generated decay lengths to obtain a reasonable 
match to distributions seen in collider data and full detector simulations. 
In fact, we find that detector modeling has relatively little effect on the 
estimated uncertainties in $\Mt$ for this method. As described in 
section~\ref{sec:tevatron}, for the Tevatron study we consider backgrounds 
such as $Wb\overline{b}$ and misidentification of light quark jets as $b$ jets 
(``mistags'') in $W$~+~jets and  multi-jet events. For the LHC, as described 
in section~\ref{sec:lhc}, we choose an event selection that reduces 
backgrounds to negligible levels. In addition, a jet veto is used to suppress events
containing significant QCD radiation, thereby minimizing the 
uncertainties associated with Monte Carlo modeling of these processes.
In section~\ref{sec:prospects} we 
discuss ways in which improvements are likely or possible. Finally, we 
summarize our results and present our conclusions in 
section~\ref{sec:conclusion}.

The results presented below, while representative of what one could 
reasonably expect to observe, include values for mean b decay lengths
in \ttbar~events that are not intended to exactly predict those
that one would extract experimentally. For example, the 
kinematic criteria used in event selection, as well as detector specific 
effects, will cause variations in the observed mean $b$ hadron decay length 
for a given top quark mass. Nevertheless, we have estimated uncertainties 
carefully and we have verified that they are robust to variations 
in the modeling of detector effects. Our projections for the uncertainties 
on the top quark mass are thus expected to be accurate predictions of 
what can be expected at the Tevatron and LHC in coming years. 


\section{\label{sec:method}The Method}

\begin{figure}
\includegraphics[width = 0.8\textwidth]{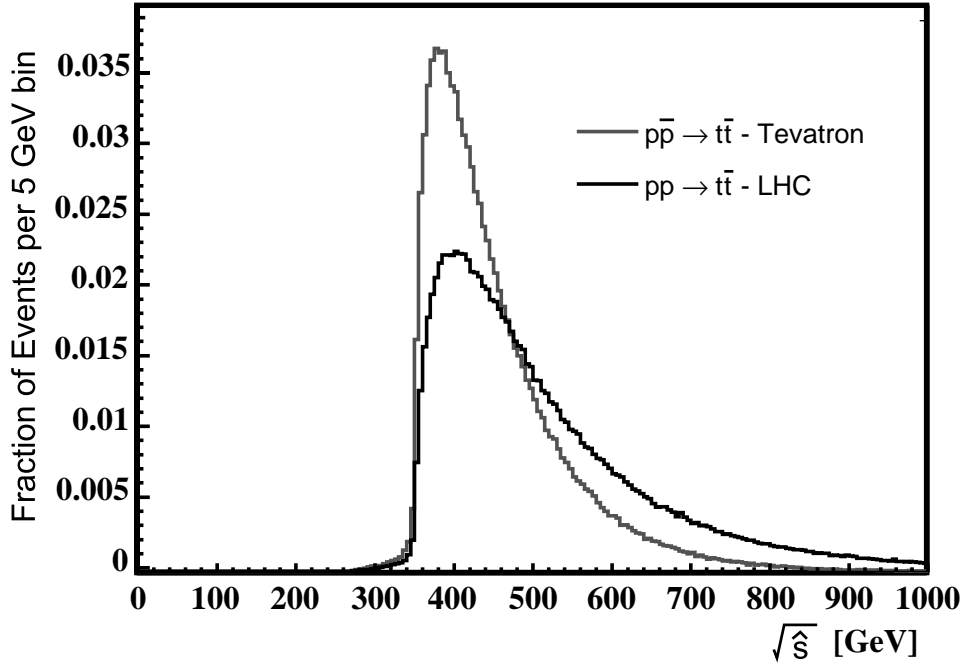}
\caption{\label{fig:cm_comp}{\tt PYTHIA} Monte Carlo comparison of the distributions of 
constituent center of mass energies ($\sqrt{\hat{s}}$) underlying \ttbar~
production at the Tevatron and LHC for $\Mt =175$ \gevcc. }
\end{figure}

The Fermilab Tevatron currently collides protons with anti-protons at 
$\sqrt{s}=1.96$~TeV. At this energy, the dominant $t\overline{t}$ production 
mechanism is \qqbar~annihilation. Valence quarks carry $\sim$15\% of the 
proton momentum on average, corresponding to a constituent collision energy of 
$\sqrt{\hat{s}}\sim300~$GeV. The \ttbar~system is therefore produced near 
threshold. The \ttbar~pair can have transverse momentum as a result of 
initial state radiation. This is typically below 100 GeV, corresponding 
to a relativistic boost that is very near to unity: $\gamma-1\le 0.04$. 
In general, 
there is little energy available to provide the individual top quarks with 
large transverse momenta. In spite of a 7-fold energy increase to 
$\sqrt{s}=14$~TeV, \ttbar~events at the LHC will be similar to those 
at the Tevatron in many ways. This is mainly due to the fact that the 
dominant production mechanism at the LHC is gluon fusion 
($gg\rightarrow t\overline{t}$). As a result, \ttbar~pairs are once again 
produced near mass threshold ($\sqrt{\hat{s}}\sim$~350-400 GeV), albeit with a 
more substantial fraction appearing at higher  
$\sqrt{\hat{s}}$. Fig.~\ref{fig:cm_comp} shows distributions  
of $\sqrt{\hat{s}}$ underlying \ttbar~production at the Tevatron and LHC 
as obtained with the {\tt PYTHIA} Monte Carlo \cite{pythia}.

In the Standard Model, the top quark decay proceeds almost uniquely as 
$t\rightarrow Wb$. In the rest frame of the top quark, the $W$ and $b$ 
daughters have equal and opposite momenta of magnitude:

\begin{equation}
p={\Mt c\over 2}\sqrt{(1-((\MW^2-\Mb^2)/\Mt^2)^2-4(\MW \Mb/\Mt^2)^2} 
\label{eq:p}
\end{equation}

Thus,  $p/c \sim 0.4 ~\Mt$ for large $\Mt$, and the relativistic boost of 
the $b$ quark is substantial,  $\gamma_b\sim 0.4\times(\Mt/\Mb)$. As a consequence, 
the $b$ quark momentum is potentially a sensitive gauge of the top quark mass.
One could use the correlation of the $b$ jet energy to the top quark mass 
directly. The drawback of this approach, however, is that it depends upon 
the $b$ jet energy measurement which suffers from a jet energy scale 
uncertainty like other methods.

As an alternative, consider the mean $b$ hadron decay length, 
$\langle L\rangle$, which is also correlated with the $b$ momentum. For a $b$ 
hadron of momentum $p$, mass $\Mb$, and proper life time $\tau_o$ one obtains: 

\begin{equation}
\langle L\rangle = c\tau_o\beta \gamma= \tau_o { p \over \Mb}  
\label{eq:l} 
\end{equation}

This expression provides the dependence on the energy since 
$E=\sqrt{(pc)^2-(\Mb c^2)^2}\sim pc$. 
The key point is that $\langle L\rangle$ 
can be found directly without measuring $E$. It is simply the average 
measured distance from the primary interaction vertex to the $b$ hadron 
decay vertex. It thus depends upon charged particle track reconstruction as opposed 
to jet energy reconstruction. Tracks are formed by fitting ``hits'' to 
helical trajectories. The hits are measurement points on charged particle 
trajectories that are typically obtained with a precision of 5-50~$\micron$ 
in semiconductor tracking detectors, such as silicon micro-strips or pixels. 

In practice, we use the average of the transverse decay length, 
$L_{xy}=L\vert \sin\theta \vert$, where $\theta$ is the angle of the $b$ 
hadron flight path with respect to the beam axis. This is necessitated by 
the fact that the net longitudinal momentum of the \ttbar~pair is not known 
in hadron collisions. The partons themselves have broad momentum
distributions within the proton or anti-proton. The sum over the transverse 
momenta of all objects in the event must however be zero. In terms of the 
transverse energy $E_T\equiv E\sin\theta$, and mass of the $b$ hadron, we have:

\begin{equation}
\langle L_{xy} \rangle  = 
c\tau_o\sqrt{ \left({\langle E_T \rangle \over \Mb c^2}\right)^2-1} 
\label{eq:lxy} 
\end{equation}

Equations \ref{eq:p} and \ref{eq:lxy} provide a simple basis for 
understanding the correlation between the mean $b$ hadron transverse decay 
length and the top quark mass. There are, however, a large number of 
additional factors that influence this correlation, and therefore need to 
be understood and taken into account. These factors can be categorized 
as follows:

\begin{itemize}
  \item Factors that affect the momentum of the top quark:
  \begin{itemize}
     \item Initial state QCD radiation from colliding partons.
     \item Final state QCD radiation from top quarks.
     \item Interference of final and initial state radiation.
     \item Parton density functions of the beam particles.
     \item Spin correlations.
  \end{itemize}
  \item Factors that affect the $b$ hadron momentum and decay length:
  \begin{itemize}
     \item Fraction of $b$ quark momentum carried by the $b$ hadron, (i.e. fragmentation).
     \item Radiation from the $b$ quark.
     \item Relative proportions of $b$ hadron species.
     \item Lifetimes of the $b$ hadron species.
     \item Modeling $b$ hadron decays.     
     \item $b$ hadron identification and vertexing efficiency 
(``$b$ tagging'').
     \item Dependence of $b$ tagging on $b$ jet momentum.
  \end{itemize}
  \item Background events mistakenly identified as \ttbar.
  \item Accidental $b$ tagging of light quark jets (``mistags'').
  \item Additional  factors:
  \begin{itemize}
     \item Kinematic selection of events.
     \item Tracking efficiency, purity, and precision.  
     \item Multiple interactions in a single beam-beam crossing.
  \end{itemize}
\end{itemize}
\subsection{\label{sec:basic} 
Basic procedure for determination of $\Mt$ and its uncertainties.}

We now give a broad overview of the basic procedures that were used in our 
studies, followed by more detailed presentations of some of the key steps. 
As mentioned above, the specific implementations of this method at the 
Tevatron and LHC differ in some respects. Details that are only 
pertinent to a specific accelerator environment are deferred to sections 
\ref{sec:tevatron} and \ref{sec:lhc}, where we describe the results
of our Tevatron and LHC studies, respectively. 

\subsubsection{\label{sec:overview} Overview } 
The procedure we follow for our studies is relatively straightforward. We 
create ``default'' \ttbar~samples using the {\tt PYTHIA} Monte Carlo. 
We do not perform a full detector simulation
of the events. We apply event selection criteria directly to the Monte Carlo generated
transverse momenta of electrons, muons, and jets in the events. 
A missing transverse energy, (\MET) requirement is applied to the transverse
component of the vector sum of the 4-vectors of all neutrinos in the event.

For a given choice of $\Mt$, we generate a large number of events. Prior 
to a determination of \Lxy~for a $b$ hadron decay, we require there to be an 
adequate number of daughter particles with significant impact parameters
relative to the primary vertex that can be used to form a secondary 
decay vertex.  The impact parameters of these particles are smeared 
according to relevant experimental resolutions. The particles must pass 
typical selection criteria. The event selection criteria and experimental
resolutions, which are typical of Tevatron and LHC experimental studies,  
are detailed in Sections~\ref{sec:tevatron} and \ref{sec:lhc}.
If it is determined that the vertex 
could potentially be observed experimentally, then the associated \Lxy~
value is obtained by simply smearing the Monte Carlo generated transverse 
decay length of the $b$ hadron.

\begin{figure}
\includegraphics[width = 0.8\textwidth]{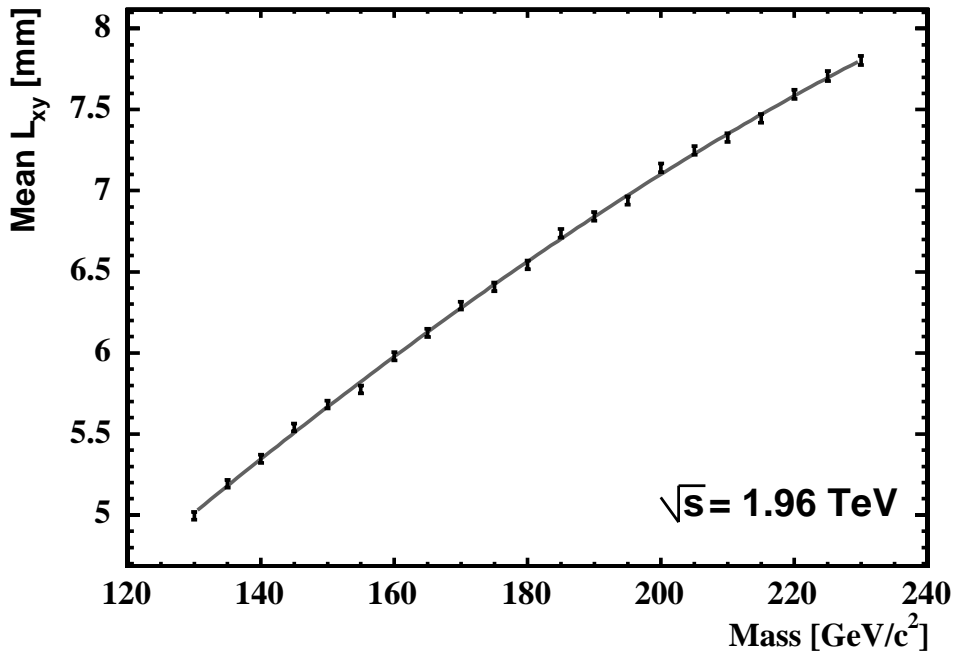}
\caption{\label{fig:lxy_smear_3to1}
An example of the correlation of \Lxy~with $\Mt$ for the
default Monte Carlo and a particular choice of event selection at the Tevatron.
Backgrounds, as described in the text, are also included
and represent 25\% of all entries.
The correlation is well fit by a third degree polynomial over the
full mass range as shown, and it is linear over small mass ranges. 
Each point represents $\sim$50,000 $b$ jets in Monte Carlo \ttbar~events.}
\end{figure}

This procedure is repeated for different choices of $\Mt$. We also generate 
background event samples that are selected and processed in a similar 
manner. We combine signal and background events in the proportions expected 
for a given experiment and choice of event selection criteria. For each 
$\Mt$, the combined signal plus background sample is used to obtain a 
mean value of the transverse decay length, \lxybar. These values are plotted 
as a function of $\Mt$ and they are fit to a polynomial in $\Mt$,
as seen in Fig. \ref{fig:lxy_smear_3to1}. For a given choice of beam energy, 
event selection, and $b$ tagging algorithm, this polynomial represents our
expectation for the correlation of $\Mt$ with \lxybar~in \ttbar~events. 
We will refer to the polynomial as the $mass$ $estimator$.

To estimate statistical uncertainties, we form an ensemble of event samples in
which the number of $b$ hadrons from signal and background events are each 
randomly selected from Poisson distributions for which mean values correspond 
to the expectations for a specific experiment and integrated luminosity.  
The \Lxy~values for a given sample in the ensemble are generated using 
the shape of the \Lxy~distribution obtained for a given choice of signal plus 
background, as described above. For each sample, \lxybar~is calculated and 
the mass estimator specific to the accelerator environment under consideration 
is used to obtain a corresponding value for $\Mt$. The resulting distribution 
of $\Mt$ values for the full ensemble is not Gaussian, in general, since the 
relationship of $\Mt$ to \lxybar~is not linear over large mass ranges 
as seen in Fig.~\ref{fig:lxy_smear_3to1}. In these cases, which
are relevant for small statistics experiments, one obtains asymmetrical
statistical uncertainties.  For high statistics experiments, only small mass 
ranges need to be considered and the resulting ensemble distribution 
for $\Mt$ is Gaussian. 
In these cases the standard deviation is taken to be the 
statistical uncertainty. The uncertainties themselves are obtained by 
constructing Neymann frequentist confidence intervals following the 
prescription described in \S 32.3.2.1 of reference \cite{Eidelman:2004wy}.

To take into account uncertainties associated with the modeling of the 
underlying physics, (e.g. initial and final state radiation, and 
parton distribution functions), we vary the relevant Monte Carlo control 
parameters to produce alternative event samples from which we can again 
extract values of \lxybar. The original mass estimator is again used to 
extract values of $\Mt$, which differ from our default values. For any 
particular variation, the difference with respect to the default, $\Delta\Mt$, 
is our estimate of the uncertainty associated with this particular aspect of
the event model. Monte Carlo parameter variations that we use are 
consistent with the current guidance provided by the authors of the 
Monte Carlos, or they conform to current conventions used in measurements of 
top quark properties at the Tevatron or studies performed by detector 
collaborations in preparation for LHC operation.

\subsection{\label{sec:b_model}\Lxy~distributions for $b$, $c$ and light
quark jets and their associated uncertainties.}
We now briefly discuss how \Lxy~is determined experimentally and the procedure used 
to obtain a reasonable approximation to \Lxy~using Monte Carlo generated information 
for $b$ and $c$ hadrons. We also discuss how we treat mistagged light quark jets. 
This is followed by a discussion of the dependence of $b$ tagging on the \ET~of 
the $b$ jet. We then present how we estimate the systematic errors on 
\lxybar~associated with the modeling of $b$ fragmentation and the 
uncertainty in the average $b$ hadron lifetime. We make the reasonable
assumption that properties of $b$ jets 
as measured in $Z\rightarrow$\bbbar~events at LEP and SLD, and
high \pt~$b$jets from direct production of \bbbar~at the Tevatron, apply
also to $b$ jets from top quark decays. This can be justified by noting 
that the final stage of jet hadronization is a non-perturbative QCD
process at a scale of order $\Lambda_{QCD}$
(see for instance \S 4.1 of \cite{Group:2004cx}).

\subsubsection{\label{sec:btag} 
Modeling \Lxy~in $b$, $c$ and light quark jets}

The experimentally observed \Lxy~distribution for $b$ and $c$ jets is the result 
of a complicated process of particle tracking, track selection criteria,
and vertex finding. While each element of this process contributes to the 
final shape of the \Lxy~distribution to some degree, the final outcome can
be described by two main effects. Namely, a convolution of the true \Lxy~
distribution with a Gaussian resolution function, and a skewing of the 
distribution that results from the use of tracks with significant impact 
parameters relative to the primary vertex. In this section we describe
in more detail how these effects arise and how we treat them with a 
Monte Carlo simulation.
 
\begin{figure}
\includegraphics[width = 0.7\textwidth]{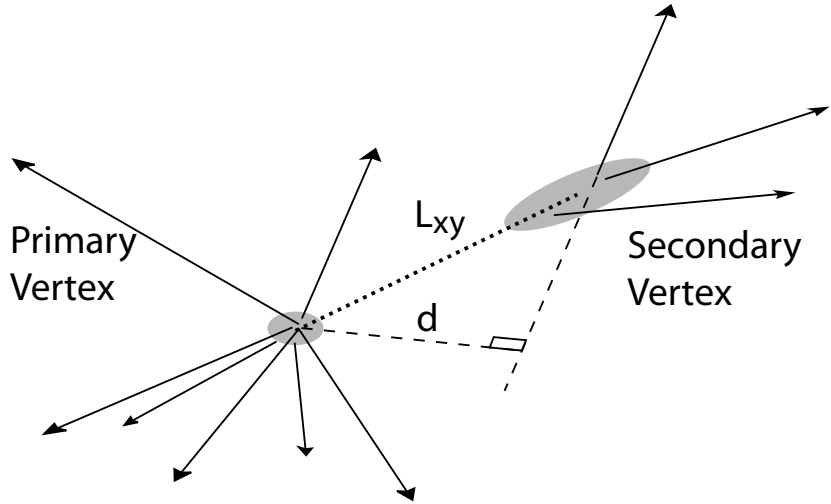}
\caption{\label{fig:tracks} Schematic representation of tracks 
in the vicinity of the beam-beam interaction for an 
event with a high energy b jet.  }
\end{figure}

The transverse decay length is the measured distance between the point at 
which the primary beam particles collide, (the primary vertex), and the 
point at which the $b$ hadron decays, (the secondary vertex), as projected 
into a plane perpendicular to the beam axis. In an actual experiment, a vertex 
is calculated as the intersection point of two or more charged particle tracks 
which have helical trajectories in the magnetic field of the tracking 
detector. Fig. \ref{fig:tracks} presents a schematic view 
of the tracking environment for a $b$ jet from top quark decay.

The primary vertex is reconstructed using all tracks that can be 
consistently associated with a specific, small region along the beam 
line. In practice, the transverse position of the primary vertex is 
sometimes taken from the beam line itself, as measured with many 
thousands of tracks, from thousands of successively recorded events, under 
a given set of operating conditions. The position of the $b$ hadron decay 
is obtained by forming a vertex from ``displaced tracks'', (i.e. tracks that
have significant impact 
parameters relative to the beam axis). Typically a track impact parameter,
$d$, is said to be significant when $S_d\equiv{d/\sigma_d}\ge x$,
where $\sigma_d$ is the uncertainty on $d$, and $x$ is typically 2.5 
or 3. The tracks are also required to be inside a cone centered on the
axis of a jet, which is defined as the line joining 
the primary vertex to the centroid of a cluster of calorimeter towers
with energies above some threshold. For high energy $b$ jets, a cone 
radius $R \equiv \sqrt{\Delta\eta^2+\Delta\phi^2} = 0.4$ is used, where 
$\Delta\eta$ and $\Delta\phi$ are the opening angles of the cone 
in pseudorapidity and azimuth, respectively \cite{geomnote}.

The sources of uncertainty in the measurement of \Lxy~
include the uncertainties in the positions of the vertices due to 
uncertainties in the reconstructed helical track parameters, which, 
in turn, are affected by hit position uncertainties and detector 
misalignments. Other sources of mis-measurement are the inclusion of fake 
tracks, and the association of tracks from other vertices. Furthermore, in 
\ttbar~events, the $b$ hadrons are not fully reconstructed and so it is often 
the case that the tracks used to locate the secondary decay vertex are a 
mixture of tracks from the $b$ hadron decay and sequential $c$ hadron 
decay. The sequential $c$ hadron has substantial momentum and in fact 
travels on average 1-2 mm beyond the $b$ hadron decay point. Use  
of such vertices is necessary to obtain high $b$ tagging efficiency but 
requires one to loosen some vertexing criteria, such as the maximum allowed 
$\chi^2$ value associated with the fit to a single vertex, but also 
increases the average uncertainty in \Lxy.
For the case of several displaced tracks from a single decay
vertex, the uncertainty on the transverse position of the vertex is
typically 100-200 $\micron$. This increases to $\sim$1 mm 
when the tracks used in the vertex come from both the $b$ and sequential 
$c$ hadron decays. 

In our studies, we use smeared Monte Carlo generated information to reproduce
the main experimental aspects of the \Lxy~distributions for $b$ and $c$ hadrons. 
For $b$ and $c$ jets we use the 4-vectors of daughter particles with
transverse momenta above 0.5 GeV/c to calculate impact parameters with 
respect to the beam line. We take into account the curvature of the tracks in
the magnetic field in this calculation. We then associate an uncertainty with
the impact parameter of each track using parameterizations that 
vary with the transverse momentum of the track as:

\begin{equation}
\sigma_d=\sqrt{\alpha^2+(\beta/p_t)^2} 
\label{eq:sigma_d}
\end{equation}

In this expression $\alpha$ represents the asymptotic impact parameter
resolution for arbitrarily high track 
momenta. The second term, which depends on $p_T$, takes into account 
multiple scattering due to interactions with the detector material. The 
parameter $\beta$ thus depends on both the amount and location of material 
in the tracker relative to where track hits are measured. The parameter 
$\alpha$ includes the uncertainty associated with the primary vertex. This can 
be quite small when the track multiplicity is large, as is often true 
for \ttbar~events. If instead one uses the position of the beam line, 
the relevant uncertainty is then the transverse size of the beam itself, 
which is $\sim 30~\micron$ at the Tevatron and 
will be $\sim 15~\micron$ at the LHC. In our studies we use 
different values for these parameters for the Tevatron and LHC. 
In all cases, we require that the tracks have an impact parameter significance 
$S_d \ge 2.5$. We only consider those $b$ and $c$ hadrons which have at 
least 2 significantly displaced tracks and we calculate an associated \Lxy~by
shifting the true transverse decay length by a random value obtained from a 
Gaussian distribution centered at the origin with standard deviation 
$\sigma_L=$~1~mm.  This value is consistent with Monte Carlo studies of 
$b$ tagging with a full simulation of the CDF detector at the Tevatron 
\cite{Acosta:2004hw}. The Gaussian smearing makes it possible for there to 
be negative decay lengths, (corresponding to the unphysical case of a secondary 
decay appearing behind the primary interaction with respect to the jet 
direction), while the requirement of at least two significantly 
displaced tracks skews the \Lxy~distribution toward positive values. The
smearing and skewing just described are the most salient features of the 
\Lxy~distribution as it is manifested experimentally \cite{Acosta:2004hw}.

We generated large numbers of events for a range of top quark masses. We also 
generated large samples of background events. For each $b$ and $c$ jet with at 
least two significantly displaced tracks, we determine \Lxy~as described 
above. We then calculated \lxybar~for combinations of top quark mass samples plus
background samples and fit the results to obtain a parameterization of 
\lxybar~versus $\Mt$. For background events we included the \Lxy~values for 
mistagged light quark jets. Note that we obtain our normalizations 
for the signal and background jets contributing to the final \Lxy~distribution 
directly from recent experimental results \cite{Acosta:2004hw}, 
scaled to the particular integrated luminosities and \ttbar~production rates 
relevant to our studies. 

In light quark jets, such as those from $W$ decay in \ttbar~events or in 
backgrounds, it is possible for significantly displaced tracks to occur and 
lead to a false $b$ tag. The probability that a light quark jet will be 
mistagged is typically less than 1$\%$, and is generally well parameterized by 
collider di-jet data as a function of jet \ET~and
pseudorapidity \cite{Acosta:2004hw}. Mistags are due to tracks with 
significant impact parameters arising from several sources. For the most 
part, they are tracks that originate at the primary interaction, but appear 
to be significantly displaced because they are on the extremes of the Gaussian
resolution distribution. True displaced 
tracks from particles with significant lifetimes (e.g. $K_s$, $\Lambda^o$, 
and photon conversions) are also sometimes incorrectly used in $b$ tagging 
when the long-lived particle is not properly identified. The latter occurs
when at least one of the daughter tracks is not properly reconstructed.
Finally, there are fake 
tracks that result from pattern recognition errors such as the association 
of hits from two or more particles in the reconstruction of a single track. 
Fake tracks and tracks from long-lived particles can result in reconstructed 
vertex locations distributed more or less uniformly in space out to or beyond the 
radius of the beam pipe or first measurement layer. We model these using a 
flat distribution at positive \Lxy.

\begin{figure}
\includegraphics[width = 0.8\textwidth]{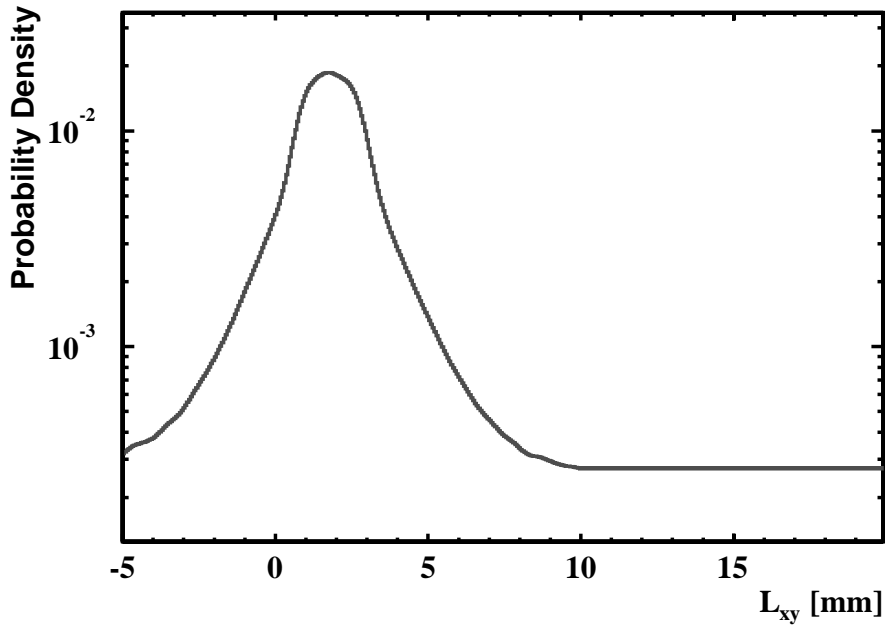}
\caption{\label{fig:mistag_model}
The probability distribution versus \Lxy~for light quark jets with at 
least two significantly displaced tracks.}
\end{figure}

Our model of the \Lxy~distribution for mistagged jets is shown in 
Fig.~\ref{fig:mistag_model}. The distribution is consistent with measurements 
and full simulation of the CDF detector \cite{Acosta:2004hw}. It is well 
approximated by the convolution of a Gaussian centered at $L_{xy}\sim$~2~mm, 
with symmetrically rising and falling exponentials in the negative and 
positive \Lxy~regions, respectively, and a flat distribution at positive 
\Lxy. As mentioned above, the distribution 
is skewed into the positive \Lxy~region as a result of the displacement 
significance requirement, $S_d\ge 2.5$, applied to the tracks.

As a check of the sensitivity of our results to our modeling of detector 
effects, we repeated the process described above in the absence of Gaussian 
smearing of \Lxy~values, and ignoring mistagged jets. We found that the  
parameterization of \lxybar~versus $\Mt$ displayed a shift in the average 
transverse displacements. The slope, however, differed negligibly 
from our previous result. The scale shift means that the two cases yield 
different mass predictions for a given value of \lxybar. The invariance of 
the slope, however, means that the estimated systematic uncertainties are the 
same in both cases. 
The reason that the slope does not change appreciably stems from the fact 
that the \Lxy~distribution for any given top quark mass is approximately an 
exponential, while the uncertainties associated with the determination of 
\Lxy~are Gaussian. The mass information is contained in the slope of
the distribution which is not changed by convolution with a Gaussian. 
The visibility of the exponential, however, is diminished as the width of the 
Gaussian increases relative to the mean of the exponential. In the case of
top quark decays, the mean decay length is roughly 5~mm while the
\Lxy~resolution is typically 1 mm or less. Thus, our
studies do not predict \lxybar~exactly for a given $\Mt$, but do provide 
robust predictions for uncertainties on $\Mt$.
 
\subsubsection{\label{sec:btag_vs_et}
Tagging efficiency versus transverse energy.}
The dependence of the $b$ tagging efficiency on jet transverse energy, \ET, 
is influenced by 
many factors including the choice of the $b$ tagging algorithm and 
the overall performance of the tracking system. For simplicity, we assume a 
dependence on jet \ET~similar to that measured with $b$ jet enriched data 
samples at the Tevatron \cite{Acosta:2004hw}. Thus, we take the efficiency 
to rise linearly in the range $15\le E_T\le 40$ GeV from a minimum of 
25\% to a maximum of 40\%. It then remains approximately constant 
for higher values of \ET. The average efficiency will affect the overall statistical 
uncertainty that can be achieved on $\Mt$ for a fixed integrated luminosity, 
while the dependence on \ET~affects \lxybar~and can therefore introduce a 
systematic uncertainty in $\Mt$. 

Studies indicate that the shape of this distribution in data is well matched 
by fully simulated Monte Carlo samples \cite{Acosta:2004hw}. However, the 
fact that the efficiency is not constant in the low \ET~region can lead to 
an uncertainty in $\Mt$ as a result of the jet energy scale uncertainty. This 
follows 
from the fact that event selection criteria usually involve thresholds 
for jets in the low \ET~region in order to have good 
signal efficiency. The jet energy scale uncertainty thus represents a
threshold uncertainty that translates into an uncertainty in the energy 
spectrum of $b$ jets contributing to the \Lxy~distribution.  For 
instance, we find that a variation in the minimum \ET~threshold by an amount 
equal to a jet energy scale uncertainty of 10\%, leads to a 1.09\% 
systematic uncertainty in $\Mt$ in our Tevatron study.

Note, however, that the $b$ tag efficiency need not be parameterized in 
terms of 
the jet \ET. As an alternative, the total \pt~of the tracks involved in the 
tag, or associated with the $b$ jet, could be used \cite{shochet}. The 
uncertainty is then shifted from the category of calorimetry to that of 
tracking, analogous to the basic premise of the correlation of $\Mt$ 
with \lxybar. In this manner, the overall uncertainty due to the variation 
of tagging efficiency with the $b$ jet \ET~could be reduced substantially.

\subsubsection{\label{fragmentation}
Fragmentation, $b$ hadron species, and their lifetimes}
The $b$ quarks from top quark decays pick up light quarks from the vacuum to create
$b$ mesons and baryons. The $b$ hadron carries a fraction $X_b$ of the initial 
$b$ quark momentum, while the remainder is dispersed into the momenta of the 
lighter mesons created in this fragmentation process. Together with the $b$ 
hadron decay products, these tracks form the $b$ jet, which is generally well 
contained within a cone of radius R = 0.4 in \ttbar~events. The fragmentation 
function, namely the distribution of $X_b$ values for many $b$ jets, has been 
studied extensively. The most precise measurements of $b$ 
fragmentation are from 
the study of $Z\rightarrow b\overline{b}$ events at the LEP and SLC 
accelerators \cite{opal,sld,aleph,delphi}. For our studies, we use the single 
most precise measurement which comes from the OPAL experiment\cite{opal}:

\begin{equation}
X_b=0.7193\pm0.0016~(stat)~^{+0.0038}_{-0.0033}~(syst).\label{eq:xb}
\end{equation}

Fragmentation is important in the correlation of $\Mt$ with \lxybar~since 
the momentum of the $b$ hadron determines the relativistic boost in 
Eq.~\ref{eq:lxy}. Thus, uncertainties in the mean, $\langle X_b\rangle$, 
and in the shape of the $X_b$ distribution could contribute 
systematic uncertainties to $\Mt$. 
Since \lxybar~is directly proportional to the average transverse momentum 
of the $b$ hadrons, we determine the uncertainty associated with 
$\langle X_b\rangle$ by simply varying \lxybar~by a fractional amount equal to
the fractional uncertainty in $\langle X_b\rangle$. Variations of the 
shape of the $X_b$ distribution could affect \lxybar, even for a fixed 
$\langle X_b\rangle$, through the processes of event selection and 
$b$ tagging. Nevertheless, since OPAL data does not strongly distinguish 
between a variety of theoretical models \cite{opal} for $b$ fragmentation in 
$Z$ decays, we do not associate any uncertainty with the shape of the 
fragmentation function. 

A potential source of uncertainty associated with our treatment of $b$ 
fragmentation stems from the fact that the color environment in \ttbar~events 
is different from that of $Z$ events. However, recent calculations 
\cite{Corcella:2004hj} indicate a value of $\langle X_b\rangle$ in top quark decays 
that is within one standard deviation of the value in Eq.~\ref{eq:xb}. We have
therefore assumed that the results of previous measurements can be applied
to \ttbar~events.

Taking the preceding discussion into account, we estimate a systematic 
uncertainty in $\Mt$ associated with $b$ fragmentation by varying
\lxybar~by $\pm$0.57\%, corresponding to the total uncertainty in
Eq.~\ref{eq:xb}. 

The fragmentation process results in the production of all $b$ hadron species. 
These appear in specific proportions, as measured by CDF and the LEP and 
SLD experiments \cite{Abbaneo:2001bv}, and with specific lifetimes 
\cite{Eidelman:2004wy}. As for $b$ fragmentation, uncertainties in the $b$ 
hadron lifetimes are important since \lxybar~depends on them directly.
To take this into account in our studies, we use the current uncertainty on 
the inclusive $b$ hadron lifetime \cite{Group:2004cx}:
\begin{equation}
\label{eq:blife}
\tau_b=1.574\pm0.008\times 10^{-12}~sec 
\end{equation}
This corresponds to a 0.51\% variation in \lxybar~if all types of $b$ hadrons 
are tracked and vertexed with equal efficiency. If this is not the case, 
then the variation may need to be amended and an uncertainty must be introduced
in association with the modeling of $b$ hadron decays.

The current values and uncertainties for the proportions and lifetimes of the 
various $b$ hadron species are summarized in Table \ref{tab:bspecies}. For 
Monte Carlo \ttbar~events with full simulation of the CDF detector and $b$ 
tagging algorithm \cite{Acosta:2004hw}, the $b$ meson species are tagged 
at a uniform rate, while the $b$ baryons tagging efficiency is 80\% of the 
meson efficiency \cite{lambdab_eff}. 
This means that the effective mean lifetime of the $b$ hadrons
entering into our \lxybar~estimate is higher than the average $b$ hadron 
lifetime in Eq.~\ref{eq:blife} by an amount 
$\Delta\tau=0.007\pm0.002\times 10^{-12}$.
This represents a $0.40\pm 0.1\%$ shift where the error takes into 
account the uncertainties in tagging and in the $b$ baryon
fraction, but does not include an uncertainty
associated with the modeling of b baryon decays, for which there exists
limited data at this time. For this reason we inflate our estimate 
by a factor of two, resulting in a
0.2\% uncertainty in \lxybar~resulting from the lower 
$b$ tagging efficiency for $b$ baryons. It is our expectation that 
Tevatron Run II studies of $b$ baryons will permit a more rigorous 
treatment of this uncertainty.

\begin{table}
\caption{\label{tab:bspecies} Fractions of $b$ hadron species in 
$Z\rightarrow b\overline{b}$ decays and their lifetimes.}
\begin{ruledtabular}
\begin{tabular}{|l|c|c|c|}
Species & Percentage & Lifetime & Relative Efficiency \\
\hline
$B^o,\overline{B^o}$ & $39.7\pm 1.0$ & $1.536\pm0.014$ & 1.0 \\
$B^{\pm}$ & $39.7\pm 1.0$ & $1.671\pm0.018$ & 1.0 \\
$B_s,\overline{B_s}$ & $10.7\pm 1.1$ & $1.461\pm0.057$ & 1.0 \\
$b~mesons$ & $90.1\pm 1.7$ & $1.603\pm0.022$ & 1.0 \\
$b~baryons$ & $9.9\pm 1.7$ & $1.208\pm0.051$ & $0.8\pm 0.05$ \\
\end{tabular}
\end{ruledtabular}
\end{table}

\subsection{\label{sec:pt_top}
Uncertainties associated with Monte Carlo modeling of \ttbar~events.}
There are a variety of uncertainties associated with the {\tt PYTHIA} Monte Carlo 
modeling of \ttbar~events that pertain to the production of jets. To 
ascertain the corresponding uncertainties in $\Mt$, we follow the direct 
advice of the authors \cite{pythiasysts} or the current conventions of the 
CDF top quark mass group to vary control parameters involved in  
\ttbar~event generation. The resulting datasets are then treated in the 
same manner as the default sample, and the resulting \lxybar~values are used 
to extract values of $\Mt$ by means of the default mass estimator. The
difference between the value obtained for each variation and that obtained 
with the default sample is then taken to be the corresponding systematic 
uncertainty.

The aspects of \ttbar~event generation that could affect jet production rates, 
in order of appearance in the Monte Carlo generation process are:
\begin{enumerate}
	\item Spin correlations.
	\item Initial state radiation.
	\item Final state radiation.
        \item Interference of initial and final state radiation. 
	\item Multiple interactions.
\end{enumerate}

Many of these are constrained by experimental results. Final state 
radiation in $W\rightarrow q\overline{q'}$ decay, for instance, is equivalent 
to that in $Z\rightarrow q\overline{q}$ decay which is very well understood 
from LEP I \cite{Bardin:1997gc}. 

Spin correlations in {\tt PYTHIA} are present for the $b$ and the $W$ decay 
products, but not for correlations between the two top quarks. In real 
experiments, there are also multiple interactions in beam-beam crossings.  
These phenomena are expected to have small effect on $\Mt$ results 
for this method and were not included in our studies.

\begin{figure}
\includegraphics[width = 0.75\textwidth]{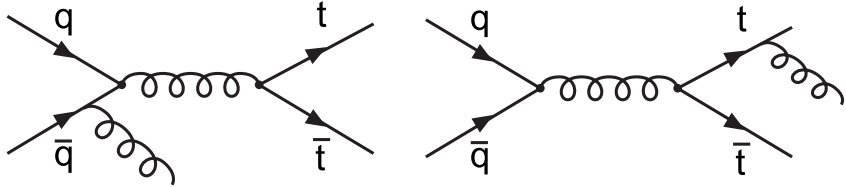}
\caption{\label{fig:isr_fsr}
Feynman diagrams for lowest order initial and final state radiation in 
\ttbar~production. Interference between these processes is also taken into 
account. Final state radiation is also possible from the $b$ quark and from 
the quarks appearing in hadronic decay of the $W$.}
\end{figure}

Lowest order Feynman diagrams for initial and final state radiation from top 
quarks are shown in Fig.~\ref{fig:isr_fsr}. To estimate the uncertainties 
associated with final state radiation we follow guidance provided by the 
authors of {\tt PYTHIA} to the CDF top group \cite{pythiasysts} in regard to 
the variation of the scale, $\Lambda_{QCD}$, and the degree of interference with 
initial state radiation. 
For initial state radiation, this guidance is again applied to our LHC study, 
while for the Tevatron, we use variations currently 
used by the CDF experiment that were determined by their studies of Drell-Yan 
di-muon events \cite{un-ki}.

With regard to final state radiation, it is thought that {\tt PYTHIA} may
underestimate the gluon emission rate relative to the \ttbar~+ jet matrix 
elements at LHC energies \cite{pythiasysts} and so one might suspect, a priori,
that the prescription used for the Tevatron could be inadequate for 
estimating LHC uncertainties. However, the event selection criteria that we 
use for the LHC, as presented below, is specifically chosen to suppress 
events with significant radiation. We therefore used the same 
parameter variations for final state radiation for both cases.

For all of our studies we use the CTEQ5 \cite{Lai:1999wy} 
parton density functions (PDFs) to create our default Monte Carlo event 
samples. To estimate the uncertainties associated with the PDFs,
we generate \ttbar~events using two variations of the MRS-98 
\cite{Martin:1998sq} PDFs for which $\Lambda_{QCD}$, 
has values of 300 GeV and 229 GeV. The difference in final \lxybar~values 
between the default CTEQ5 sample and the MRS-98 sample with $\Lambda=300$~GeV 
is added in quadrature with the difference in the two MRS-98 samples 
resulting from the variation of $\Lambda_{QCD}$. This procedure, while  
$ad$ $hoc$ in appearance, follows historical conventions at hadron colliders. 
Recently, however, a more sophisticated alternative has been suggested by the 
CTEQ collaboration \cite{Pumplin:2002vw} which we will apply 
in section~\ref{sec:prospects}. 

To estimate an uncertainty on the modeling of the top quark \pt~spectrum at 
the Tevatron, we took the tree-level spectrum obtained from {\tt PYTHIA} and 
re-weighted the events to match the NNLO spectrum \cite{Kidonakis:2004hr} 
at 1.96 TeV, normalized to equal area. We found the resultant shift in the 
extracted top quark mass to be 0.9 \gevcc~for a generated mass of $\Mt=178$ \gevcc.
We draw several conclusions from this. First, for the purpose of 
our studies, the difference is small enough to justify our use of {\tt PYTHIA} 
without inclusion of higher order matrix elements. Second, the difference 
between the tree-level and NNLO spectra is expected to be significantly
larger than the uncertainty on the NNLO calculation alone. Hence, in an 
actual measurement of $\Mt$, one would avoid a significant uncertainty
by using a Monte Carlo for which the top quark \pt~spectrum is 
consistent with the NNLO prediction. 
Furthermore, at the LHC the \ttbar~production rates 
are large enough to allow one to 
define an unbiased sample from which one could do a direct top quark \pt~spectrum 
measurement. The Monte Carlo could be compared to measurement and tuned 
accordingly to allow the associated uncertainty in $\Mt$ to be reduced to a 
negligible level. We therefore do not assign an uncertainty associated 
with the top quark \pt~spectrum in our studies.

\section{\label{sec:results}
Estimates based upon currently available information.}
 
As seen in Section~\ref{sec:method}, many of
the systematic uncertainties associated with this method are largely determined
by the precision of experimentally determined quantities, such as the
average $b$ hadron lifetime.  
In this section we present estimates for the uncertainty on $\Mt$ for the Tevatron 
and the LHC using only information that is available
at the present time. This has the virtue of yielding very sound estimates
of systematic uncertainties. It does not, however, give a true picture of the 
precision that might be achieved with this method in the future, when much of
the information that the method relies upon will be more refined that it is 
at present. This is particularly true in regard to the application of this method 
at the LHC where direct measurements with LHC \ttbar~data will allow several of
the larger systematic uncertainties to be reduced. This is
discussed in Section~\ref{sec:prospects} below.

\begin{figure}
\includegraphics[width = 0.9\textwidth]{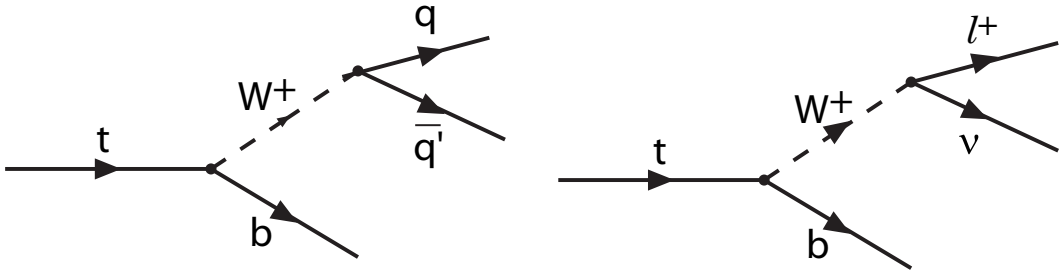}
\caption{\label{fig:signatures}
The final state signatures for \ttbar~events are determined by the
$W$, which can decay either to quarks or to leptons. To determine
$\Mt$ at the Tevatron, we studied the lepton + jets signature for \ttbar~
events in which one $W$ decayed leptonically and the other hadronically. 
For the LHC we studied the dilepton signature for \ttbar~in which the $W$'s 
decay to an electron or muon and their respective neutrino.}
\end{figure}

As mentioned in section \ref{sec:introduction}, 
the differences between the two environments lead to different event selection 
criteria. We focus on the lepton~+~jets event signature
for the Tevatron, and dilepton event signature for the LHC, (see 
Fig. \ref{fig:signatures}). In both cases, 
the term lepton actually refers to either a muon or electron from $W$ decay 
and not a tau lepton, which is most often manifested experimentally as a 
narrow hadronic jet. Events in which the $W$ decays leptonically tend to 
have smaller backgrounds, and hence dilepton event samples, in which both
$W$'s decay leptonically, are the most pure. The all-hadronic \ttbar~events
are those for which both $W$'s decay to quarks.  The dilepton, lepton+jets, 
and all-hadronic \ttbar~events represent roughly 5\%, 30\%, and 45\% of all 
\ttbar~events, respectively. 

At the Tevatron, where \ttbar~yields are much smaller than at the LHC, 
the lepton+jets
category provides the most significance. Our Tevatron study, therefore, is
focused on this signature. We do, however,  estimate 
the extent to which the uncertainty in $\Mt$ can be reduced by 
inclusion of dilepton events.
For Tevatron event selection criteria, backgrounds contribute an uncertainty to 
$\Mt$ as a result of the uncertainties associated with both the normalization and 
shape of the \Lxy~distribution for each of the various background types. 

At the LHC, it is more important to
minimize systematic uncertainties. 
Thus, we focus on the dilepton signature to suppress backgrounds and
reduce to minimize the occurrence of QCD radiation from final state partons.  
For the same purpose, we normalize our event counts to be consistent with
the requirements that both $b$ jets are tagged, and that there are 
no other jets in the event. 
As mentioned earlier, the ``jet veto'' also helps to minimize the probability that
there is significant QCD radiation in the event.

\subsection{\label{sec:tevatron}Estimates for the Fermilab Tevatron}

Run II of the Tevatron is currently in progress. At the time of this writing, 
data from an integrated luminosity $\sim$0.5~$\fb$ have 
been accumulated by each of 
the two collider experiments, CDF and D0. By the end of 
Run II it is expected that this will reach a value of 4.5 to 8.5 $\fb$. 

As mentioned above, the use of the lepton+jets signature for \ttbar~events 
is a compromise between sample purity and \ttbar~event counts. 
Uncertainties associated with the modeling and normalizations of 
backgrounds contribute a Tevatron-specific uncertainty in $\Mt$. We thus
start with a discussion of our treatment of these backgrounds
before summarizing all of our projected systematic and 
statistical uncertainties.
\subsubsection{\label{sec:tev_selection}Tevatron event selection.}
%

For our modeling of \ttbar~events at the Tevatron, the tracks used to 
determine whether a given $b$ or $c$ hadron decay could produce an observable 
displaced vertex, as discussed in section \ref{sec:btag}, have 
impact parameter resolution comparable to that of the CDF
experiment \cite{Acosta:2004hw}. Thus, in Eq.~\ref{eq:sigma_d}, we use
parameters: $\alpha=$~36 $\micron$ and $\beta=$~25~$\micron$-GeV/c. 
Events are selected by requiring one lepton and one 
neutrino with transverse momentum of at least 20 GeV/c. There must be a 
total of 3 or more well-separated partons with transverse momenta of at 
least 15 GeV/c. 

\subsubsection{\label{sec:background}Tevatron backgrounds.}
%
We considered various processes that can have similar characteristics to 
\ttbar~events, and thus represent background constituents of 
Tevatron data sets. The relative proportion of any particular background type
depends strongly on the event selection. We employ
selection criteria used by the CDF collaboration for the 
lepton~+~jets signature \cite{Acosta:2004hw}, and therefore also assume the 
same types and relative proportions of backgrounds, with the exception of 
$WW$, $WZ$, $ZZ$ and $Z\rightarrow\tau\tau$ events which together represent only 
2.3\% of all backgrounds and so were ignored for simplicity.  
We thus consider the following final states: $W$\bbbar, $W$\ccbar, $Wc$, 
\bbbar~+ jets, and mistags in $W$+jet events. Background sources
are listed in Table~\ref{tab:backgrounds}.

\begin{figure}
\includegraphics[width = 0.8\textwidth]{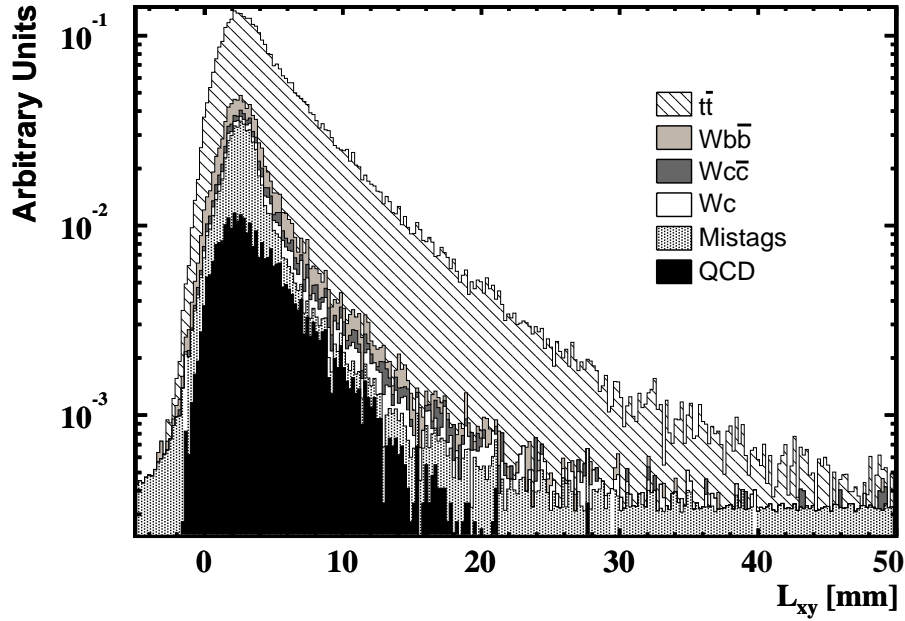}
\caption{\label{fig:mt175_smear}
The \Lxy~distribution for \ttbar~with $\Mt=175$ \gevcc, and the \Lxy~
distributions for various backgrounds in lepton + jets events at the 
Tevatron. Default Monte Carlo parameters were used in addition to the 
parametric detector resolution model discussed in the text. The signal to 
background ratio is 3 to 1.}
\end{figure}

The contributions from single top quark production were also investigated. 
The t-channel production of a single top quark involves a 
rather different \pt~spectrum 
than that found in \ttbar~events and could therefore be treated as a 
``background''. However, it makes a negligible contribution 
after our event selection criteria are applied. The s-channel production of a
single top quark has essentially the same dependence of \lxybar~on $\Mt$ as that 
of \ttbar~events and thus represents a potential signal source, but turns out
to be negligible in comparison to the contribution from \ttbar. 
For simplicity, we do not include either process in our studies.

\begin{table}
\caption{\label{tab:backgrounds} Various sources and proportions of 
backgrounds used in our studies, following ref. \cite{Acosta:2004hw}.}
\begin{ruledtabular}
\begin{tabular}{|l|c|} 
Source & Fraction \\
\hline
$W$\bbbar~& 20.5\% \\
$W$\ccbar~& 7.7\% \\
$Wc$ & 8.9\% \\
\bbbar~+ jets & 30.4\% \\
$W$+jets (mistag) & 32.5\% \\
\end{tabular}
\end{ruledtabular}
\end{table}

Backgrounds were simulated using {\tt ALPGEN} matrix elements 
\cite{Mangano:2002ea}, (including up to 2 additional radiated partons), 
together with the {\tt HERWIG} parton shower model \cite{herwig}. Each sample was 
processed with the same selection criteria as that applied to the \ttbar~signal. 
For backgrounds involving charm, the $c$ hadron daughter 
tracks were used for the tagging simulation, and the kinematic cuts 
previously applied to $b$ jets were applied to the $c$ jets. 

The signal for a given $\Mt$ was combined with the total background 
in a 3-to-1 ratio. The \Lxy~distributions for \ttbar, ($\Mt=175$ \gevcc), 
and backgrounds is seen in Fig.~\ref{fig:mt175_smear}. 
The resulting values of \lxybar~for the various
$\Mt$ choices were then used to define a mass estimator for the Tevatron
study. 

\subsubsection{\label{sec:Tevatron_results}Tevatron Results.}
%

\begin{figure}
\includegraphics[width = 0.8\textwidth]{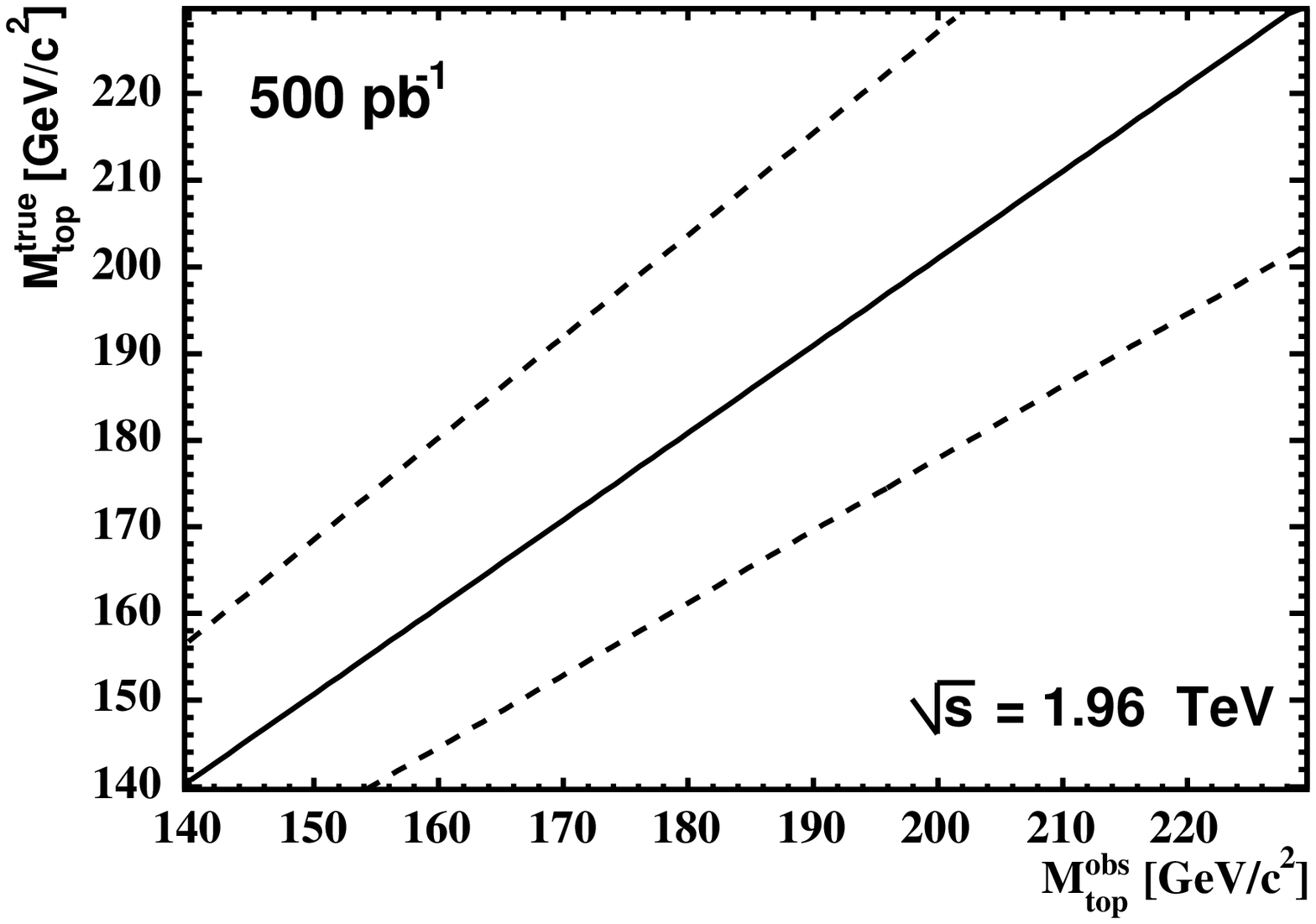}
\caption{\label{fig:mtrue_tev_500pb}
The correlation of the input, (``true''), value of $\Mt$ with the measured, 
(``observed''), value of $\Mt$ (solid) is plotted for ensembles of Monte 
Carlo lepton~+~jets events for the Tevatron. The dashed lines are the one 
standard deviation contours (statistical only) for 500 $\pb$ integrated 
luminosity. Default Monte Carlo parameters were used in addition to the 
parametric detector resolution model discussed in the text. The signal to 
background ratio is 3 to 1.}
\end{figure}

\begin{figure}
\includegraphics[width = 0.8\textwidth]{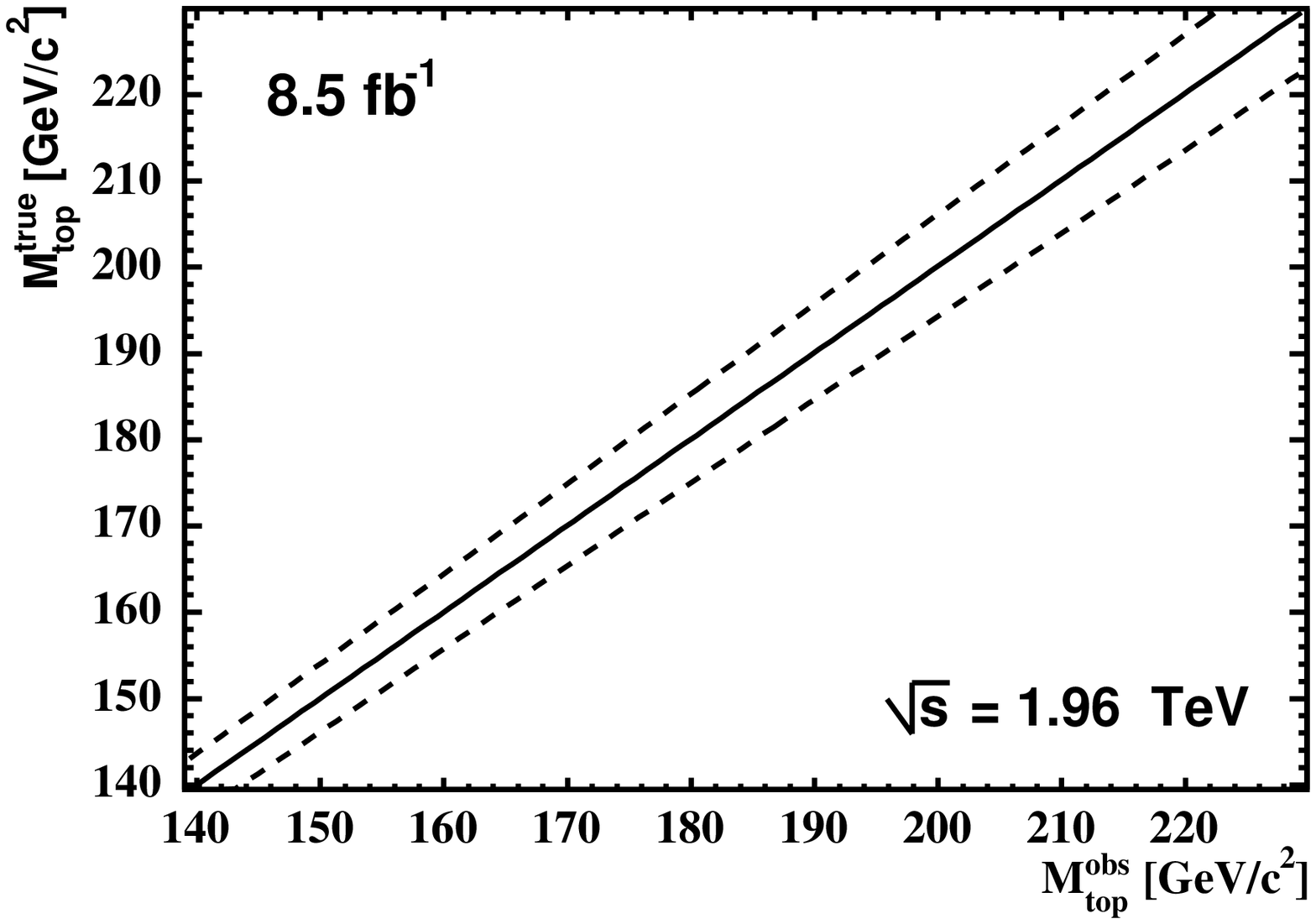}
\caption{\label{fig:mtrue_tev_max}
The correlation of the input, (``true''), value of $\Mt$ with the measured, 
(``observed''), value of $\Mt$ (solid) is plotted for ensembles of Monte 
Carlo lepton~+~jets events for the Tevatron. The dashed lines are the one 
standard deviation contours (statistical only) for 8.5 $\fb$ integrated 
luminosity. Default Monte Carlo parameters were used in addition to the 
parametric detector resolution model discussed in the text. The signal to 
background ratio is 3 to 1.}
\end{figure}

We determine the statistical uncertainty as described in 
section~\ref{sec:method} for two different integrated luminosities at the 
Tevatron: the current value of 500~$\pb$ and the maximum projected Run II value 
of 8.5 $\fb$. For each case, we estimate the corresponding numbers of top quarks
and background events that we would expect to have by scaling results
recently reported by CDF \cite{Acosta:2004hw}. For top quark masses in the
range 130~$\le\Mt\le$~230~\gevcc, we
create ensembles of large numbers of Monte Carlo samples as described
in section~\ref{sec:method}, and determine \lxybar~for each. The mass 
estimator is then used to convert each value of \lxybar~to a value of $\Mt$. 
The distribution of ``observed'' values of $\Mt$ for a given ensemble has 
average value consistent with the mass value input to the Monte Carlo.
From the distribution of observed $\Mt$ values for a given input mass 
and choice of integrated luminosity, we determine statistical uncertainties 
by constructing Neymann frequentist confidence intervals. 
Figures~\ref{fig:mtrue_tev_500pb} and \ref{fig:mtrue_tev_max} show the 
results for integrated luminosities of 500 $\pb$ and 8.5 $\fb$, respectively.
In these figures, the dashed lines represent the confidence intervals
used to determine statistical uncertainties. Thus, for instance, for a top quark 
mass of $\Mt = 178$~\gevcc,  we expect that measurements done at the Tevatron 
will have statistical uncertainties of 20.8~\gevcc~and 5.0~\gevcc~
for integrated luminosities of 500~$\pb$ and 8.5~$\fb$, respectively. 

The systematic uncertainties, obtained by the methods described in
section~\ref{sec:method}, are the same for both cases and are listed in 
Table~\ref{tab:tevatron}. The statistical uncertainty on the individual 
systematics listed in the table is  $\sim$0.2~\gevcc.
\begin{table}
\caption{\label{tab:tevatron} Systematic uncertainties for
$\Mt=175$~\gevcc~for the Tevatron.}
\begin{ruledtabular}
\begin{tabular}{|l|c|}
Source 	& Uncertainty \gevcc \\
\hline
Initial state radiation  & 0.7	\\
Final state radiation & 1.1 \\
Parton distributions & 0.8 \\
$b$ hadron lifetime & 1.0 \\
$b$ fragmentation  & 0.9 \\
$b$ tagging & 0.8 \\
Backgrounds & 0.3	\\
Jet Energy Scale & 1.9	\\ 
\hline
TOTAL & 2.9 \\
\end{tabular}
\end{ruledtabular}
\end{table}

To estimate the effect of adding \ttbar~dilepton events, we scale event 
counts and tagged $b$ jets based upon the results from a recent CDF 
\ttbar~dilepton 
cross section measurement \cite{Acosta:2004uw}. For an integrated luminosity 
of 8.5 $\fb$ the statistical error is reduced from 5.0 to 4.4 \gevcc. For the 
most part, the systematic uncertainties associated with the dilepton signature 
will be similar to those for the lepton+jets signature, except for the uncertainty
due to backgrounds which will be smaller. 


\subsection{\label{sec:lhc}Estimates for the CERN LHC}

\begin{figure}
\includegraphics[width = 0.8\textwidth]{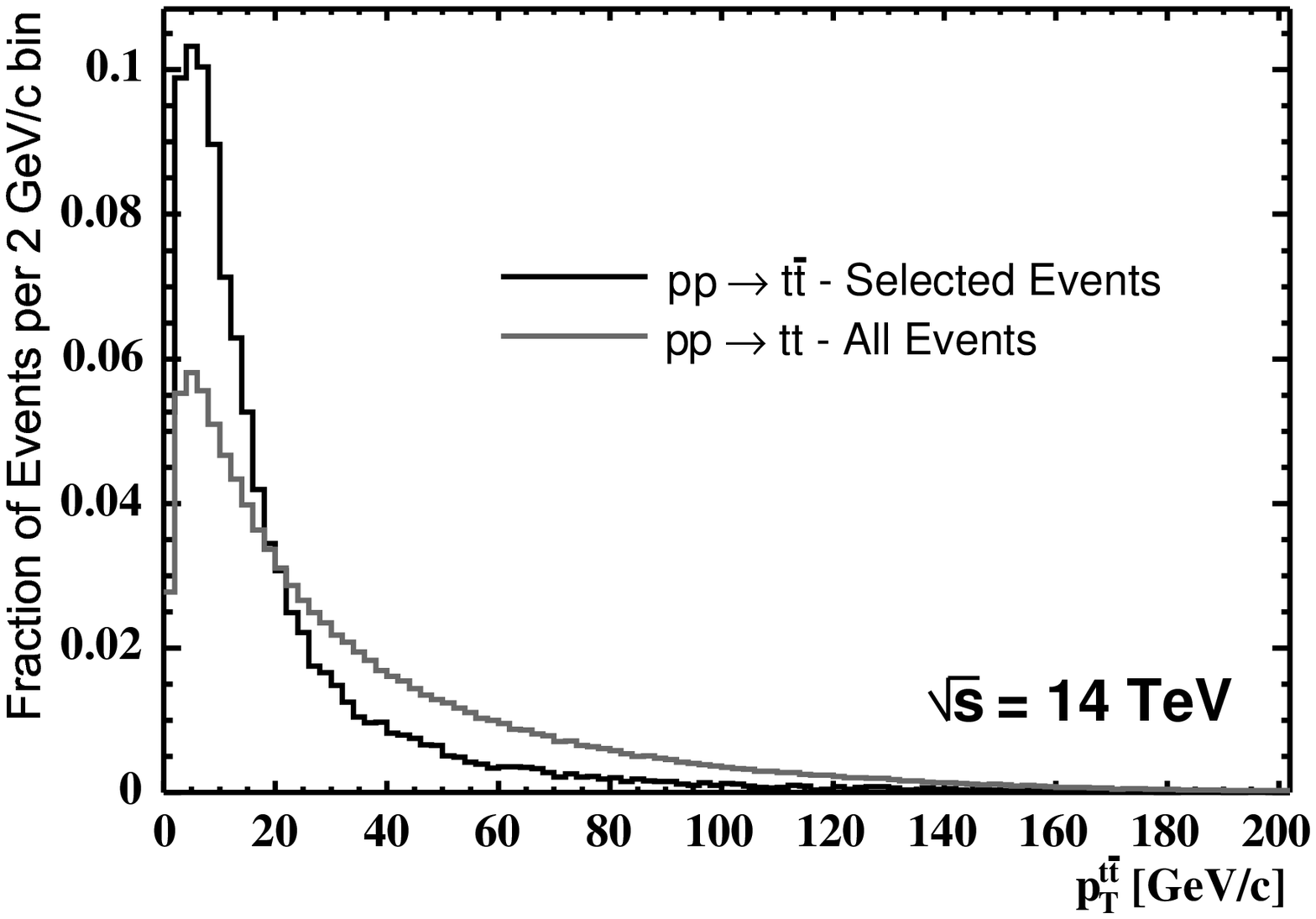}
\caption{\label{fig:ttbar_pt_lhc}
The transverse momentum of the \ttbar~pair at $\sqrt{s}=14$~TeV is plotted
for all events, and for only those events passing the dilepton
event selection discussed in the text. The curves are normalized
to unit area.}
\end{figure}

For our LHC study, anticipated \ttbar~production rates allow us to define
event selection criteria that minimize uncertainties. We select 
only dilepton events in which the two leptons have transverse momenta of at 
least 35 and 25 GeV/c. We also require two $b$ jets with 
transverse energy above 15 GeV/c and no other clustered energy above 
10 GeV. These event selection criteria effectively eliminate all backgrounds while
minimizing the likelihood of QCD radiation. The effect of event selection on the
transverse momenta of the top quarks is seen in Fig~\ref{fig:ttbar_pt_lhc}. 
For an integrated luminosity of 10~$\fb$, these requirements yield an 
expectation of $\sim$~13,000 events, out of a $\sim$ 8,000,000 
produced \ttbar~events. It is estimated that these events have a ratio of 
\ttbar~signal to backgrounds in excess of 30-to-1 
\cite{karel}. We therefore do not include an uncertainty associated with
backgrounds in our LHC study. 

For our modeling of \ttbar~events at the LHC, we use track impact parameter 
resolution parameters comparable to those of the CMS experiment 
\cite{cmstdr}: $\alpha=$~15 $\micron$ and $\beta=$~75 $\micron$~-~GeV/c.  

\begin{figure}
\includegraphics[width = 0.8\textwidth]{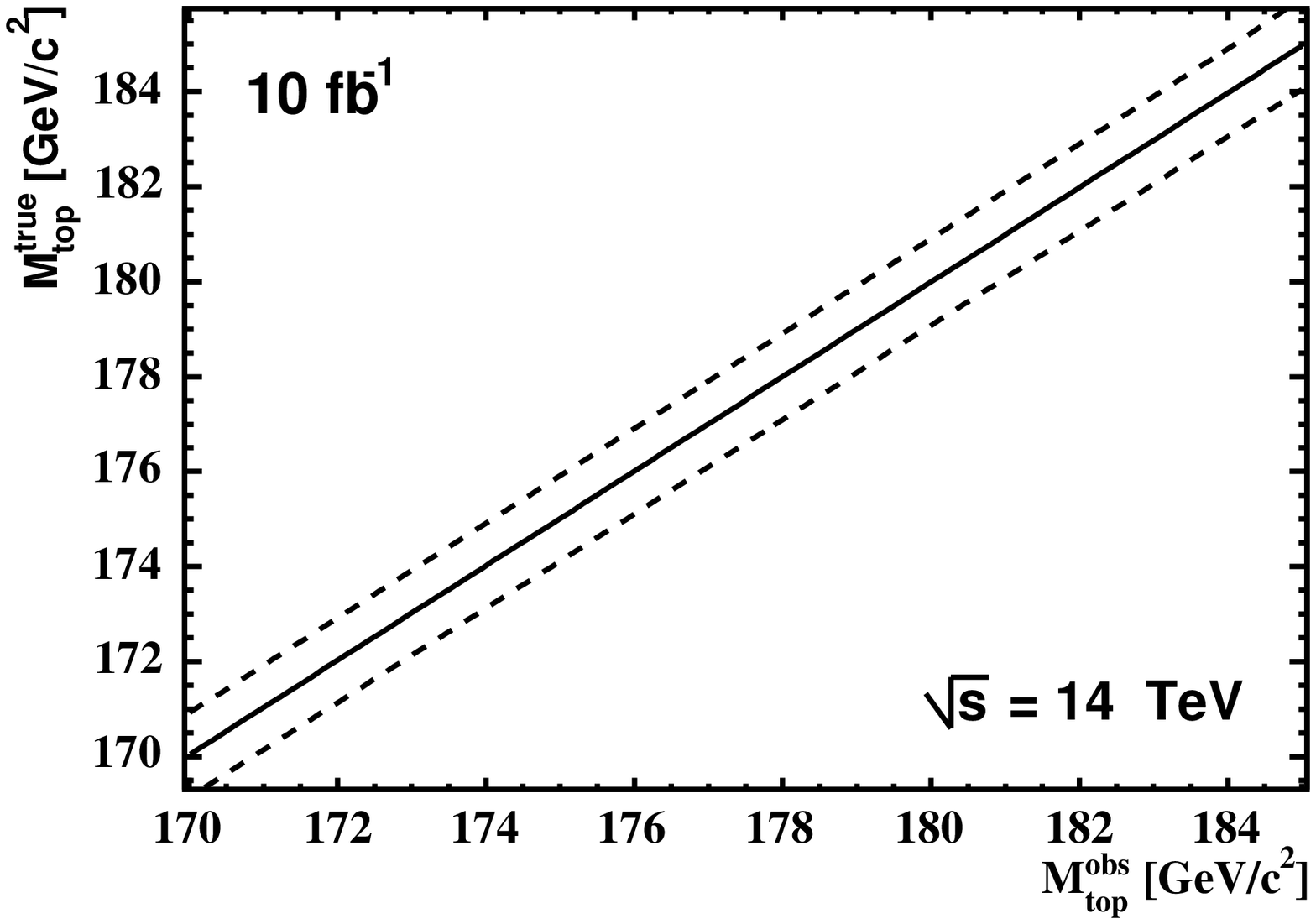}
\caption{\label{fig:mtrue_lhc_10fb}
The correlation of the input, (``true''), value of $\Mt$ with the measured, 
(``observed''), value of $\Mt$ (solid) is plotted for ensembles of Monte 
Carlo dilepton events for the LHC. The dashed lines are the one standard 
deviation contours (statistical only) for 10 $\fb$ integrated luminosity.}
\end{figure}

We estimate a statistical uncertainty of 0.9 \gevcc~for the LHC in 10 $\fb$,
as seen in Fig~\ref{fig:mtrue_lhc_10fb}. The systematic uncertainties are 
listed in Table~\ref{tab:lhc}. As for the Tevatron, the statistical 
uncertainty on the individual entries in the table is $\sim$0.2~\gevcc.

\begin{table}
\caption{\label{tab:lhc} 
Systematic uncertainties for $\Mt=175$ \gevcc~for the LHC study.}
\begin{ruledtabular}
\begin{tabular}{|l|c|}
Source  & Uncertainty \gevcc \\
\hline
Initial state radiation  & 1.3	\\
Final state radiation & 0.5 \\
Parton distributions & 0.7 \\
$b$ hadron lifetime & 1.3 \\
$b$ fragmentation  & 1.2 \\
Jet Energy Scale & 0.2	\\ 
\hline
TOTAL & 2.4 \\
\end{tabular}
\end{ruledtabular}
\end{table}

We did not include an uncertainty associated with $b$ tagging efficiency. 
It is our expectation that the large \ttbar~samples available at the LHC 
will allow the $b$ tagging efficiency to be measured as a function of the summed
\pt~of the tracks, or the \ET~of the $b$ jet, with small uncertainty.
To estimate the effect of the jet energy scale uncertainty, we varied jet energies by
$\pm$3\% in accordance with the CMS hadron calorimeter technical design
report \cite{hcal:1997pw}.

\begin{figure}
\includegraphics[width = 0.8\textwidth]{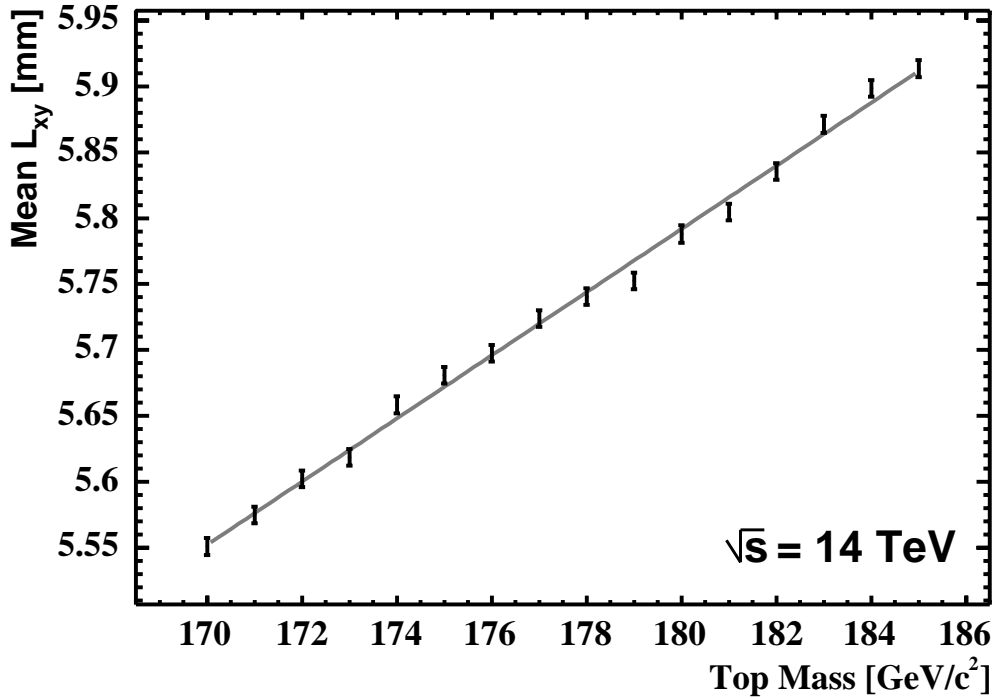}
\caption{\label{fig:lxy_vs_mtop_lhc}
An example of the correlation of \Lxy~with $\Mt$ for the
default Monte Carlo and our particular choice of event selection criteria for 
the LHC. The correlation is well fit by a straight line over small mass ranges 
such as the one pictured here. Each point represents $\sim$ 26,000 b jets 
in \ttbar~events.}
\end{figure}

\section{\label{sec:prospects} Future improvements. }

In the preceding sections we presented our estimates for the uncertainties 
on $\Mt$ using the mean decay length of $b$ hadrons in \ttbar~events at the 
Tevatron and LHC. We used only currently available 
information for modeling of \ttbar~
events. For the most optimistic Tevatron Run II integrated luminosity 
scenario, and using events with the 
lepton~+~jets signature, we obtained statistical and systematic uncertainties 
of 5.0 and 2.9 \gevcc, respectively. We estimated that the statistical 
uncertainty can be reduced to 4.4 \gevcc~by inclusion of events conforming 
to the dilepton signature, without significantly affecting the systematic 
uncertainty. This corresponds to a total uncertainty of $\sim$5 \gevcc. 
Since our estimated statistical uncertainty is larger than our estimated
systematic uncertainty, we do not expect this result to be improved without
an increase in total integrated luminosity or an increase in the event
identification or b tagging efficiencies. With regard to the former
some small improvement could be obtained by loosening event selection
criteria and allowing background to rise until an optimum total uncertainty
is attained. With regard to $b$ tagging, we have used currently achieved
efficiencies, (in the range of 25-40\%), that could turn out to be  
conservative.
For instance, the CDF experiment has only recently begun to include
information from the Layer 00 silicon detector 
\cite{Hill:2003ax,Nelson:2001dg}, 
which provides track hits at an average radius 1.5 cm that leads to a
substantial improvement in track impact parameter resolution. The D0 collaboration 
is now constructing a similar small radius silicon detector 
\cite{Hanagaki:2003cn} to enhance their 
$b$ tagging. These devices, as well as gradual improvements in the  
understanding and alignment of tracking detectors at the Tevatron will 
likely result in higher $b$ tagging efficiencies. Thus, the actual
statistical uncertainties on $\Mt$ per Tevatron experiment, for any particular
integrated luminosity, could be lower than the estimates 
presented above.

For 10 $\fb$ integrated luminosity at the LHC, we projected statistical and 
systematic uncertainties of 0.9 and 2.4 \gevcc, respectively, corresponding 
to a total uncertainty of 2.5 \gevcc. Such precision is already significant,
as it would allow for an important cross-check or improvement of other 
measurements of $\Mt$. There is, however, the possibility for reduction 
of several of the systematic uncertainties associated with our LHC study.

As seen in Table~\ref{tab:lhc}, the dominant systematic uncertainties at the 
LHC are those associated with
initial state radiation, $b$ fragmentation, and the average $b$ hadron lifetime.
The uncertainty in the average $b$ hadron lifetime, and 
lifetimes of the individual $b$ hadron species, continue to improve with
time and the accumulation of significantly larger \bbbar~event samples
at $B$ factories and the Tevatron. It is therefore  
reasonable to anticipate that by the time this measurement is performed at the 
LHC, there will be a reduction in the uncertainties
associated with the $b$ lifetimes. 
The uncertainties associated with QCD radiation will be 
significantly improved as a result of direct measurements at the Tevatron and LHC.
Event samples containing millions of \ttbar~events at the LHC will 
allow stringent constraints to be placed on the 
Monte Carlo modeling of QCD radiation that could reduce the corresponding
systematic uncertainties to negligible levels. This in turn will allow event
selection criteria to be loosened considerably to reduce the statistical
uncertainty without significant penalty in the systematic uncertainty on $\Mt$.

On the other hand, there are other systematic uncertainties that may not 
improve.  For example, the average momentum carried
by the $b$ hadron in the process of fragmentation is an 
experimentally determined quantity that is currently dominated by the 
measurements with $Z\rightarrow b\overline{b}$ events at LEP and SLD, 
which are limited by systematic 
uncertainties \cite{Eidelman:2004wy}. These measurements 
also benefited a great deal from 
the beam-energy constraint provided by collisions of electrons with 
positions and are thus unlikely to be significantly improved at hadron colliders. 
It may, however, be possible for some overall improvement to be obtained 
by combining existing measurements. 

With regard to parton distribution functions, it is difficult for us to say 
how much LHC data might improve the estimates we presented in 
Section~\ref{sec:lhc}, although some
improvement seems inevitable. However, as we noted  
in Section~\ref{sec:pt_top}, a more sophisticated algorithm 
PDF uncertainties has recently been developed. The CTEQ collaboration has 
established a new framework that enables  
correlations between the experimental measurements used to 
define the PDFs to be taken into account.  It provides a pragmatic way to quantify 
uncertainties via an eigenvector approach to the Hessian method
\cite{Pumplin:2002vw}. The eigenvectors form an orthonormal basis
for a 20 dimensional parameter space of the 20 free parameters
of the CTEQ6M fit. Forty-one PDF sets are generated, corresponding 
to positive and negative excursions along each of the 20 orthogonal axes in 
the parameter space, and the central CTEQ6M set. These can be used to estimate
the uncertainty for any physical quantity that depends upon them, including
the top quark mass, by adding in quadrature the differences between the 
positive and negative sets for each of the eigenvectors. When we apply
this procedure to our Tevatron analysis, we obtain an uncertainty that is
$\sim$15\% smaller than our previous result shown in Table~\ref{tab:tevatron}. 

These considerations lead us to conclude that by the
time a measurement of the top quark mass has been performed 
at the LHC with 10 $\fb$ of integrated luminosity, the result will be 
substantially more precise than the estimate we presented in section 
\ref{sec:lhc}. For example, a factor of two reduction in the uncertainties
associated with QCD radiation, together with smaller 10-40\% reductions
in the uncertainties associated with PDFs, $b$ fragmentation, and average
$b$ hadron lifetime, reduce our total estimated uncertainty to $\sim 1.5$ 
\gevcc~per LHC experiment.

\section{\label{sec:conclusion}Conclusion}

We have presented a new method for the measurement of the 
top quark mass at the Tevatron and LHC that is largely uncorrelated with other
methods. Initial studies using currently available information to
model all aspects of \ttbar~events yields uncertainties on $\Mt$ of
$\sim$5 \gevcc~for 8.5 $\fb$ of integrated luminosity in Run II of the Tevatron
and $\sim$2.5 \gevcc~for 10 $\fb$ of integrated 
luminosity at the LHC. Furthermore, with likely improvements
in our understanding of $b$ hadron properties and \ttbar~events 
the uncertainties associated with this method at the LHC
could be substantially reduced, making this method comparable in mass
resolution to all other methods, for which a total uncertainty
of 1.5-2.0 \gevcc~per experiment is expected.

\begin{acknowledgments}
We'd like to acknowledge the following people: D. Barge, 
C. Campagnari, F. Gianotti, D. Glenzinski, D. Rainwater,
K. Smolek and U.K. Yang for insightful comments and contributions to 
this paper. 
This work was made possible by U.S. Department of Energy grant DE-FG03-91ER40618.
\end{acknowledgments}
\bibliography{lxy_prd}

\begin{thebibliography}{42}
\bibitem[{lep()}]{lepton}
\bibinfo{note}{Another quantity that does not suffer from jet energy scale
  uncertainty that we considered is the mean transverse energy of leptons from
  $W$ decays in \ttbar~events. Initial studies indicate that the correlation is
  too weak to be useful at the Tevatron, but could become important at the LHC.
  This was first suggested to us by C. Campagnari, The University of California
  Santa Barbara.}
\bibitem[{geo()}]{geomnote}
\bibinfo{note}{We use a cylindrical coordinate system about the beam axis in
  which $\theta$ is the polar angle, $\phi$ is the azimuthal angle, and $\eta
  \equiv -\ln \tan (\theta /2)$. $E_T \equiv E \sin \theta$ and $p_T \equiv p
  \sin \theta$ where $E$ is energy measured by the calorimeter and $p$ is
  momentum measured by the spectrometer. \MET$\equiv -\sum_i E_T^i {\bf n}_i$,
  where ${\bf n}_i$ is the unit vector in the azimuthal plane that points from
  the beam line to the $i$th calorimeter tower.}
\bibitem[{sho()}]{shochet}
\bibinfo{note}{This was first suggested to us by M. Shochet, University of
  Chicago.}
\end{thebibliography}
\end{document}